\documentstyle[sprocl]{article}
\makeatletter
\def\@cite#1#2{{[{#1}]\if@tempswa\typeout
{IJCGA warning: optional citation argument
ignored: `#2'} \fi}}

\newcount\@tempcntc
\def\@citex[#1]#2{\if@filesw\immediate\write\@auxout{\string\citation{#2}}\fi
  \@tempcnta\z@\@tempcntb\m@ne\def\@citea{}\@cite{\@for\@citeb:=#2\do
    {\@ifundefined
       {b@\@citeb}{\@citeo\@tempcntb\m@ne\@citea\def\@citea{,}{\bf ?}\@warning
       {Citation `\@citeb' on page \thepage \space undefined}}%
    {\setbox\z@\hbox{\global\@tempcntc0\csname b@\@citeb\endcsname\relax}%
     \ifnum\@tempcntc=\z@ \@citeo\@tempcntb\m@ne
       \@citea\def\@citea{,}\hbox{\csname b@\@citeb\endcsname}%
     \else
      \advance\@tempcntb\@ne
      \ifnum\@tempcntb=\@tempcntc
      \else\advance\@tempcntb\m@ne\@citeo
      \@tempcnta\@tempcntc\@tempcntb\@tempcntc\fi\fi}}\@citeo}{#1}}
\def\@citeo{\ifnum\@tempcnta>\@tempcntb\else\@citea\def\@citea{,}%
  \ifnum\@tempcnta=\@tempcntb\the\@tempcnta\else
   {\advance\@tempcnta\@ne\ifnum\@tempcnta=\@tempcntb \else \def\@citea{--}\fi
    \advance\@tempcnta\m@ne\the\@tempcnta\@citea\the\@tempcntb}\fi\fi}
\makeatother

\def\Ref#1{Ref.~\cite{#1}}

\def\gtino{\wt g_{3/2}}

\def\sq{\wt q}

\def\slep{\wt \ell}

\def\mslep{m_{\slep}}
\def\slepl{\wt \ell_L}

\def\slepr{\wt \ell_R}

\def\sel{\wt e}

\def\etmiss{\slash E_T}

\def\mhalf{m_{1/2}}
\def\gl{\wt g}
\def\mgl{m_{\gl}}

\def\stopone{\wt t_1}

\def\sbot{\wt b}

\def\sq{\wt q}

\def\slep{\wt \ell}

\def\mslep{m_{\slep}}
\def\slepl{\wt \ell_L}

\def\slepr{\wt \ell_R}

\def\etc{{\it etc.}}

\def\sign{{\rm sign}}

\def\eg{{\it e.g.}}
\def\etal{{\it et~al.}}
\def\mhalf{m_{1/2}}
\def\gl{\wt g}
\def\mgl{m_{\gl}}

\def\stopone{\wt t_1}

\def\sq{\wt q}

\def\slep{\wt \ell}

\def\mslep{m_{\slep}}
\def\slepl{\wt \ell_L}

\def\slepr{\wt \ell_R}

\def\sbot{\wt b}

\def\hl{h^0}
\def\hh{H^0}
\def\ha{A^0}

\def\tanb{\tan\beta}

\def\cnone{\wt\chi^0_1}
\def\cnonestar{\wt\chi_1^{0\star}}
\def\cntwo{\wt\chi^0_2}

\def\snu{\wt\nu}
\def\snul{\wt\nu_L}

\def\msnu{m_{\snu}}
\def\mcnone{m_{\cnone}}
\def\mcntwo{m_{\cntwo}}

\def\wt{\widetilde}
\def\cpone{\wt \chi^+_1}
\def\cmone{\wt \chi^-_1}
\def\cpmone{\wt \chi^{\pm}_1}

\def\mcpmone{m_{\cpmone}}

\def\staur{\wt \tau_R}

\def\mstaur{m_{\staur}}

\def\MPL #1 #2 #3 {{\sl Mod.~Phys.~Lett.}~{\bf#1} (#3) #2}
\def\NPB #1 #2 #3 {{\sl Nucl.~Phys.}~{\bf B#1} (#3) #2}
\def\PLB #1 #2 #3 {{\sl Phys.~Lett.}~{\bf B#1} (#3) #2}
\def\PR #1 #2 #3 {{\sl Phys.~Rep.}~{\bf#1} (#3) #2}
\def\PRD #1 #2 #3 {{\sl Phys.~Rev.}~{\bf D#1} (#3) #2}
\def\PRL #1 #2 #3 {{\sl Phys.~Rev.~Lett.}~{\bf#1} (#3) #2}
\def\RMP #1 #2 #3 {{\sl Rev.~Mod.~Phys.}~{\bf#1} (#3) #2}
\def\ZPC #1 #2 #3 {{\sl Z.~Phys.}~{\bf C#1} (#3) #2}
\def\IJMP #1 #2 #3 {{\sl Int.~J.~Mod.~Phys.}~{\bf#1} (#3) #2}

\def\wtil{\widetilde}

\def\br{\rm BR}

\def\gam{\gamma}

\def\etal{{\it et al.}}

\def\anti{\overline}
\def\epem{e^+e^-}

\def\mupmum{\mu^+\mu^-}

\def\rts{\sqrt s}
\def\eg{{\it e.g.}}

\def\anti{\overline}

\def\tanb{\tan\beta}
\def\hl{h^0}

\def\ha{A^0}

\def\hh{H^0}

\def\fbi{~{\rm fb}^{-1}}

\def\gev{~{\rm GeV}}
\def\tev{~{\rm TeV}}

\def\beq{\begin{equation}}
\def\eeq{\end{equation}}
\def\beqa{\begin{eqnarray}}
\def\eeqa{\end{eqnarray}}
\def\lsim{\mathrel{\raise.3ex\hbox{$<$\kern-.75em\lower1ex\hbox{$\sim$}}}}
\def\gsim{\mathrel{\raise.3ex\hbox{$>$\kern-.75em\lower1ex\hbox{$\sim$}}}}

\def\Eq#1{Eq.~(\ref{#1})}
\def\Ref#1{Ref.~\cite{#1}}

\def\ie{{\it i.e.}}
\def\hl{h^0}
\def\hh{H^0}
\def\ha{A^0}

\def\tanb{\tan\beta}

\def\wt{\widetilde}
\def\wh{\widehat}

\def\gev{~{\rm GeV}}

\def\etmiss{E_T^{\rm miss}}

\def\china{\widetilde\chi^0_1}
\def\chinb{\widetilde\chi^0_2}

\def\chipa{\widetilde\chi^+_1}

\def\ifmath#1{\relax\ifmmode #1\else $#1$\fi}

\def\fivethirds{{\textstyle{5 \over 3}}}

\def\sq{\widetilde q}

\def\gl{\widetilde g}

\def\ls#1{\ifmath{_{\lower1.5pt\hbox{$\scriptstyle #1$}}}}
\def\anti{\overline}
\def\bold#1{\setbox0=\hbox{$#1$}%
     \kern-.025em\copy0\kern-\wd0
     \kern.05em\copy0\kern-\wd0
     \kern-.025em\raise.0433em\box0 }

\begin{document}

%
\noindent
\rightline{SCIPP 97/37}
\rightline{UCD-97-29}
\rightline{December 1997}
\rightline{hep-ph/9806330}
\vspace{2.5cm}

\thispagestyle{empty}
\centerline{{\Large\bf Low-Energy Supersymmetry at Future Colliders}}
\vskip1pc
\begin{center}
{\large John F. Gunion}
\vskip3pt
{\it Davis Institute for High Energy Physics,   \\
Department of Physics, \\  University of California, Davis, CA 95616}
\vskip6pt
{\large Howard E. Haber}\\
\vskip3pt
{\it Santa Cruz Institute for Particle Physics, \\
University of California, Santa Cruz, CA 95064}
\end{center}

\vskip1cm
\centerline{\bf Abstract}
\vskip6pt
We classify the variety of low-energy supersymmetric signatures that
can be probed at future colliders.
We focus on phenomena associated with the minimal supersymmetric
extension of the Standard Model.  The structure of the
supersymmetry-breaking introduces additional model assumptions.  The
approaches considered here are supergravity-mediated and
gauge-mediated supersymmetry-breaking.
Alternative phenomenologies arising in non-minimal and/or
R-parity-violating approaches are also briefly examined.

\vfill
\begin{center}
To appear in 
{\it Perspectives on Supersymmetry}, \\
Gordon L. Kane, editor  (World Scientific, 1998)
\end{center}
\clearpage       
\setcounter{page}{1}
\title{LOW-ENERGY SUPERSYMMETRY AT FUTURE COLLIDERS
}
\author{JOHN F. GUNION}
\address{Davis Institute for High Energy Physics,   \\
Department of Physics, \\  University of California, Davis, CA 95616}
\vskip5pt
\author{HOWARD E.  HABER}
\address{Santa Cruz Institute for Particle Physics, \\
University of California, Santa Cruz, CA 95064}

\maketitle\abstracts{
We classify the variety of low-energy supersymmetric signatures that
can be probed at future colliders.
We focus on phenomena associated with the minimal supersymmetric
extension of the Standard Model.  The structure of the
supersymmetry-breaking introduces additional model assumptions.  The
approaches considered here are supergravity-mediated and
gauge-mediated supersymmetry-breaking.
Alternative phenomenologies arising in non-minimal and/or
R-parity-violating approaches are also briefly examined.
}

\section{Introduction}

In this chapter, we focus on the signatures
for low-energy supersymmetry at future colliders in the context of the
minimal supersymmetric extension of the Standard
Model (MSSM) \cite{Haber85,smartin}.
In its most general form (with the assumption of R-parity conservation), the
MSSM is a 124-parameter theory \cite{sutter,habersusy97};
most of the parameter freedom is
associated with the supersymmetry-breaking sector of the
model~\footnote{The notation for the supersymmetric parameters used in
this paper for the most part follows that of \Ref{smartin}.
The notation for supersymmetric particle names follows that of
\Ref{Haber85}.}.
This huge parameter space can be reduced by: (i) imposing
phenomenological constraints, and (ii) imposing theoretical
assumptions on the structure of supersymmetry-breaking.
In addition, the scale of supersymmetry-breaking,
$\sqrt F$, must be specified. It
determines the properties of the gravitino, $\gtino$.
In previous chapters,
two broad model categories for supersymmetry-breaking were discussed,
gravity-mediated supersymmetry breaking (SUGRA)
and gauge-mediated supersymmetry breaking (GMSB).

In most SUGRA models \cite{sugra}, $\sqrt F$ is so large that the
$\gtino$ interactions are too weak for it to play any role
in collider phenomenology.
In the {\it minimal} supergravity (mSUGRA) framework,
the soft-super\-symmetry-breaking
parameters at the Planck scale take a particularly
simple form and depend on essentially five new parameters.
These include $m_0$
(a flavor universal soft supersymmetry-breaking scalar mass),
$m_{1/2}$ (a universal gaugino mass), and $A_0$ (a flavor universal
tri-linear scalar interaction).  In particular,
gaugino mass unification
implies that at the unification scale ($M_X$), the U(1),
SU(2) and SU(3) gaugino Majorana mass parameters are equal, {\it i.e.},
$M_1(M_X) = M_2(M_X) = M_3(M_X) = m_{1/2}$.  This implies that the
{\it low-energy} gaugino mass parameters satisfy:
\beq \label{gauginomassrelation}
 M_3 = {g^2_3\over g_2^2} M_2\simeq 3.5M_2,
\qquad M_1 = \fivethirds\tan^2\theta_W M_2\simeq 0.5M_2\,.
\eeq
The other two mSUGRA parameters are the supersymmetric Higgs mass
parameter $\mu$ and an off-diagonal soft Higgs squared-mass.
After the imposition of electroweak symmetry breaking, these two
parameters can be traded in for the two Higgs vacuum expectation
values (modulo a sign ambiguity in $\mu$).  By fixing the $Z$
mass, the remaining mSUGRA parameters are determined by
the ratio of Higgs vacuum expectation values ($\tanb$) and the sign of
$\mu$.  The lightest
supersymmetric particle (LSP) is nearly always the lightest neutralino
(denoted in this chapter by $\china$).
Non-minimal extensions of the mSUGRA model have also been considered
in which some of the parameter universality assumptions have been
relaxed.

In GMSB models \cite{giudice}, $\sqrt F$ is suffiently small that the $\gtino$
is almost always the lightest supersymmetric particle (LSP)
and plays a prominent phenomenological
role.  Then, different choices for the next-to-lightest supersymmetric
particle (NLSP) lead to different phenomenologies.
In the simplest GMSB models, the gaugino and scalar
soft-supersymmetry-breaking masses
are given by SU(3), SU(2) and U(1) gauge group factors times
an overall scale $\Lambda$, while the $A$ parameters are expected to
be negligible.  [The low-energy values of the gaugino mass parameters
also satisfy Eq.~(\ref{gauginomassrelation}).]
The parameter set is then completed by
$\tanb$ and $\sign(\mu)$.

Finally, one can also consider alternative low-energy supersymmetric
approaches.  For example, if R-parity violation (RPV) is
present \cite{dreiner},
additional supersymmetric parameters are introduced.  These include
parameters
$\lambda_L$, $\lambda^\prime_L$ and $\lambda_B$ which govern new
lepton and baryon number violating
scalar-fermion Yukawa couplings derived from the following
supersymmetic interactions:
\beq
(\lambda_L)_{pmn} \widehat L_p \widehat L_m \widehat E^c_n
+ (\lambda_L^\prime)_{pmn}\widehat L_p \widehat Q_m\widehat D^c_n
+(\lambda_B)_{pmn}\widehat U^c_p \widehat D^c_m \widehat D^c_n\,,
\label{rpv}
\eeq
where $p$, $m$, and $n$ are generation indices, and
gauge group indices are suppressed.  In the notation
above, the ``superfields'' $\wh Q$, $\wh U^c$, $\wh D^c$, $\wh L$, and
$\wh E^c$ respectively represent
$(u, d)_L$, $u^c_L$, $d^c_L$, $(\nu$, $e^-)_L$, and $e^c_L$ and the
corresponding superpartners.  The Yukawa interactions are obtained
from \Eq{rpv} by
taking all possible combinations involving two fermions
and one scalar superpartner.  

\section{Classes of Supersymmetric Signals}

The lack of knowledge of the origin and structure of the
supersymmetry-breaking parameters implies that the predictions for low-energy
supersymmetry and the consequent phenomenology
depend on a plethora of unknown parameters.
Nevertheless, we can broadly
classify supersymmetric signals at future colliders by considering
a variety of theoretical approaches.  In this section, we delineate the
possible
supersymmetric signatures, and in the next section we explore their
consequences for experimentation at future colliders.

\subsection{Missing energy signatures}

In R-parity-conserving low-energy supersymmetry, supersymmetric
particles are produced in pairs.  The subsequent decay of a heavy
supersymmetric particle generally proceeds via a multistep decay chain
\cite{leveille,chain,enhancedbs}, ending in the production of at least
one supersymmetric particle that (in conventional models)
is weakly interacting and escapes the
collider detector.  Thus, supersymmetric particle production yields
events that contain at least two escaping non-interacting particles,
leading to a missing energy signature.  At hadron colliders, it is only
possible to detect missing transverse energy ($\etmiss$), since
the center-of-mass energy of the hard collision is not known on an
event-by-event basis.

In conventional SUGRA-based models, the weakly-interacting LSP's that
escape the collider detector (which yields large missing transverse
energy) are accompanied by energetic jets and/or
leptons.  This is the ``smoking-gun'' signature of low-energy supersymmetry.
However, there are two unconventional approaches in
which the smoking-gun signature is absent. First, consider a model in
which the $\china$ is the LSP but the lightest neutralino and chargino
are nearly degenerate in mass.  If the mass difference is $\lsim
100$~MeV, then $\chipa$ is long-lived and decays outside the
detector~\cite{guniondrees1,guniondrees2}. In this case, some supersymmetric
events would yield {\it no} missing energy and two semi-stable charged
particles that pass through the detector.  Second, there are models in
which a gluino (more precisely, the $R^0=\gl g$ bound state)
is the LSP~\footnote{Farrar has advocated the
existence of a very light gluino with a mass less than a few GeV
\cite{glennys}.  Recent experimental data \cite{nogluinos}
show no evidence for a such a light gluino, although
the assertion that light gluinos are
definitively ruled out is still in dispute.
The possibility of a more massive LSP gluino in SUGRA-based models has
been considered in \Ref{guniondrees2}.}.
A massive $R^0$ is likely to
simply pass through the detector without depositing significant energy.
Even when it is light enough to be stopped,
the hadronic calorimeter will measure only the kinetic energy of the $R^0$.
In either case, there would be substantial missing
energy~\cite{gunbarcelona}.  However, there would be no jets arising
from $\gl$ decays in such models.

In conventional GMSB models with a gravitino-LSP~\footnote{It is
also possible
to construct a GMSB scenario in which the $\gl$ is the LSP \cite{raby}.
The resulting phenomenology corresponds to that of the massive gluino
LSP discussed above.}, all supersymmetric
events contain at least two NLSP's, and the resulting signature depends
on the NLSP properties.  Four physically distinct possible scenarios
emerge:
\begin{itemize}
\item
The NLSP is electrically and color neutral and long-lived, and decays
outside of the detector to its associated Standard Model partner and the
gravitino.
\item
The NLSP is the sneutrino and decays invisibly
into $\nu \gtino$ either inside or outside the detector.
\end{itemize}
In either of these two cases, the resulting missing-energy signal is
similar to that of the SUGRA-based models where $\china$ or
$\snu$ is the LSP.
\begin{itemize}
\item
The NLSP is the $\china$ and decays inside the detector to $N \gtino$, where
$N=\gamma$, $Z$ or a neutral Higgs boson.
\end{itemize}
In this case, the gravitino-LSP behaves like the neutralino or sneutrino LSP of
the SUGRA-based models.  However, in contrast to SUGRA-based models,
the missing energy events of the GMSB-based model are
characterized by the associated production of (at least) two $N$'s, one for
each NLSP~\footnote{If the decay of the NLSP is not prompt, it is possible to
produce events in which one NLSP decays inside the detector and one
NLSP decays
outside of the detector.}.  Note that if $\china$ is lighter than the $Z$ and
$\hl$ then BR$(\china\to\gamma \gtino)=100\%$, and all supersymmetric
production will result in missing energy events with at least two associated
photons.
\begin{itemize}
\item
The NLSP is a charged slepton (typically $\staur$ in GMSB models
if $\mstaur<\mcnone$),
which decays to the corresponding lepton partner and gravitino.
\end{itemize}
If the decay is prompt, then one finds missing energy events with
associated leptons (taus).  If the decay is not prompt, one observes a
long-lived heavy semi-stable charged particle with {\it no} associated
missing energy (prior to the decay of the NLSP).

There are also GMSB scenarios in which there are several nearly degenerate
so-called co-NLSP's \cite{conlsp}, any one of which can
be produced at the penultimate step of the supersymmetric decay
chain~\footnote{For example, if
$\wt\tau_R^\pm$ and $\china$ are nearly degenerate in mass, then neither
$\wt\tau_R^\pm\to\tau^\pm\china$ nor $\china\to\wt\tau_R^\pm\tau^\mp$
are kinematically allowed decays.  In this case, $\wt\tau_R^\pm$ and
$\china$ are co-NLSP's, and each decays dominantly into its Standard
Model superpartner plus a gravitino.}.
The resulting supersymmetric signals would consist of events with
two (or more) co-NLSP's, each one of which would decay according to
one of the four scenarios delineated above.  For additional details on
the phenomenology of the co-NLSP's, see  \Ref{conlsp}.

In R-parity violating SUGRA-based models the LSP is unstable.  If the
RPV-couplings are sufficiently weak, then the LSP will decay outside the
detector, and the standard missing energy signal applies.  If the LSP
decays inside the detector, the phenomenology of RPV models depends on
the identity of the LSP and the branching ratio of possible final state
decay products. If the latter includes a neutrino, then the
corresponding RPV supersymmetric events would result in missing energy
(through neutrino emission) in association with hadron jets and/or
leptons.  Other possibilities include decays into charged leptons
in association with jets (with no neutrinos), and decays into purely
hadronic final states.  Clearly, these latter events would contain
little missing energy. If R-parity violation is present in GMSB
models, the RPV decays of the NLSP can easily dominate over the NLSP
decay to the gravitino. In this case, the phenomenology of the NLSP
resembles that of the LSP of SUGRA-based RPV models.

\subsection{Lepton ($e$, $\mu$ and $\tau$) signatures}

Once supersymmetric particles are produced at colliders, they do not
necessarily decay to the LSP (or NLSP) in one step.  The resulting decay
chains can be complex, with a number of steps from the initial decay to
the final state~\cite{chain}.  Along the way, decays can produce real or
virtual $W$'s, $Z$'s, charginos, neutralinos  and sleptons,
which then can produce leptons in
their subsequent decays.  Thus, many models yield large numbers of
supersymmetric events characterized by one or more leptons in
association with missing energy, with or without hadronic jets.

One signature of particular note is events containing like-sign
di-leptons~\cite{likesign}.
The origin of such events is associated with the Majorana nature of the
gaugino.  For example, $\gl\gl$ production followed by
$\gl\to q\anti q\cpmone\to q\anti q\ell^\pm\nu\cnone$
can result in like-sign leptons since the $\gl$ decay leads with equal
probability to either $\ell^+$ or $\ell^-$.
If the masses and mass differences are both substantial
(which is typical in mSUGRA models, for example), like-sign di-lepton
events will be characterized by fairly energetic jets and isolated
leptons and by large $\etmiss$ from the LSP's.
Other like-sign di-lepton signatures can arise in a similar way from the decay
chains initiated by the heavier neutralinos.

Distinctive tri-lepton signals~\cite{bcpttri} can result from
$\cpmone\cntwo\to(\ell^{\pm}\nu\cnone)(\ell^+\ell^-\cnone)$.
Such events have little hadronic activity (apart
from initial state radiation of jets off
the annihilating quarks at hadron
colliders).  These events can have a variety of interesting characteristics
depending on the fate of the final state neutralinos.

If the soft-supersymmetry breaking slepton masses are
flavor universal at the high energy scale $M_X$
(as in mSUGRA models) and $\tanb\gg 1$, then
the $\staur$ will be significantly lighter than the other slepton states.
As a result, supersymmetric decay chains involving (s)leptons will
favor $\staur$ production, leading to a predominance of events with
multiple $\tau$-leptons in the final state.

In GMSB models with a charged slepton NLSP, the decay
$\slep\to\ell\,\gtino$ (if prompt) yields at least two leptons for every
supersymmetric event in association with missing energy.  In
particular, in models with a $\staur$ NLSP, supersymmetric events will
characteristically contain at least two $\tau$'s.

In RPV models, decays of the LSP
(in SUGRA models) or NLSP (in GMSB models)
mediated by RPV-interactions proportional to $\lambda_L$ and $\lambda^\prime_L$
will also yield supersymmetric events containing charged leptons. However,
if the only significant RPV-interaction is the one proportional to
$\lambda^\prime_L$, then such events would
{\it not} contain missing energy (in contrast to the GMSB signature
described above).

\subsection{$b$-quark signatures}

The phenomenology of gluinos and squarks depends critically on their
relative masses.  If the gluino is heavier, it will decay dominantly
into $q\sq$~\footnote{In this section, we employ the notation $q\sq$ to
mean either $q\anti{\sq}$ or $\anti{q}\sq$.}, while the squark can decay
into quark plus chargino or neutralino.  If the squark is heavier, it
will decay dominantly into a quark plus gluino, while the gluino will
decay into the three-body modes $q\bar q\wt\chi$ (where $\wt\chi$ can be
either a neutralino or chargino, depending on the charge of the final
state quarks).  A number of special cases can arise when the possible
mass splitting among squarks of different flavors is taken into account.
For example, models of supersymmetric mass spectra have been considered
where the third generation squarks are lighter than the squarks of the
first two generations.  If the gluino is lighter than the latter but
heavier than the former, then the only open gluino two-body decay mode
could be $b\sbot$~\footnote{Although one top-squark mass-eigenstate
($\stopone$) is typically lighter than $\sbot$ in models, the heavy
top-quark mass may result in a kinematically forbidden gluino decay mode
into $t\stopone$.}.  In such a case, all $\gl\gl$ events will result in
at least four $b$-quarks in the final state (in associated with the
usual missing energy signal, if appropriate).
More generally, due to
the flavor independence of the strong interactions, one expects
three-body gluino decays into $b$-quarks in at least 20\% of all gluino
decays~\footnote{Here we assume the approximate degeneracy of the first
two generations of squarks, as suggested from the absence of
flavor-changing neutral-current
decays.  In many models, the $b$-squarks tend to be of similar mass or
lighter than the squarks of the first two generations.}.  Additional
$b$-quarks can arise from both top-quark and top-squark decays, and from
neutral Higgs bosons produced somewhere in the chain decays~\cite{bquarkbaer}.
Finally, at large $\tanb$, the enhanced Yukawa coupling to $b$-quarks
can increase the rate of $b$-quark production in
neutralino and chargino decays occurring at some step in the gluino
chain decay.

These observations suggest that many supersymmetric events at hadron
colliders will be characterized by $b$-jets in association with missing
energy~\cite{enhancedbs,bquarkian,snowtheory2}.

\subsection{Signatures involving photons}

In mSUGRA models, most supersymmetric events do not contain isolated
energetic photons.  However, some areas of low-energy supersymmetric
parameter space do exist in which final state photons can arise in the
decay chains of supersymmetric particles.
If one relaxes the condition of gaugino mass unification, then
the {\it low-energy} gaugino mass parameters no longer must satisfy
Eq.~(\ref{gauginomassrelation}).  As a result, interesting alternative
supersymmetric phenomenologies can arise.  For example, if
the low-energy mass parameters satisfy $M_1\simeq
M_2$, then the branching ratio for $\chinb\to\china\gamma$ can be
significant \cite{wyler}.  In the model of \Ref{eegamgamkane}, the $\china$-LSP
is dominantly higgsino, while $\chinb$ is dominantly gaugino.  Thus,
many supersymmetric decay chains end in the production of $\chinb$,
which then decays to $\china\gamma$.  In this picture, the pair
production of supersymmetric particles often yields two photons plus
associated missing energy.  At LEP-2, one can also produce
$\china\chinb$ which would then yield single photon events in
association with large missing energy.

In GMSB models with a $\china$-NLSP, all supersymmetric decay chains would end
up with the production of $\china$.  Assuming that $\china$ decays inside the
collider detector, one possible decay mode is $\china\to\gamma \gtino$.
In many models, the branching ratio for this
radiative decay is significant (and could be as
high as 100\% if other possible two-body decay modes are not kinematically
allowed).  In the latter case, supersymmetric pair production would also yield
events with two photons in associated with large missing energy.  The
characteristics of these events differ in detail from those of
the corresponding events expected in the model of \Ref{eegamgamkane}.

\subsection{Kinks and long-lived heavy particles}

In most SUGRA-based models, all supersymmetric particles in the decay chain
decay promptly until the LSP is reached.  The LSP is exactly stable and
escapes the collider detector.  However, exceptions are possible.
In particular, if there is a supersymmetric
particle that is just barely heavier
than the LSP, then its (three-body) decay rate
to the LSP will be significantly suppressed and it could be long lived.  For
example, in the models with $|\mu|\gg M_1> M_2$
\cite{guniondrees1,guniondrees2}
implying $\mcpmone\simeq\mcnone$, the $\cpmone$ can be sufficiently long
lived to yield a detectable vertex, or perhaps even exit the detector.

In GMSB models, the NLSP may be long-lived, depending on its mass and
the scale of supersymmetry breaking, $\sqrt{F}$.  The NLSP is unstable
and eventually decays to the gravitino.  For example, in the case of the
$\china$-NLSP (which is dominated by its U(1)-gaugino component), one
finds $\Gamma(\cnone\to
\gam\gtino)= m^5_{\tilde\chi_1^0}\cos^2\theta_W/16\pi F^2$.  It then
follows that
\begin{equation}
(c\tau)_{\cnone\to \gam \gtino}\simeq 130
\left({100\gev\over\mcnone}\right)^5
\left({\sqrt F\over 100\tev}\right)^4\mu {\rm m}\,.
\label{ctauform}
\end{equation}

For simplicity, assume that $\china\to\gamma\gtino$ is the dominant NLSP
decay mode.  If $\sqrt F\sim 10^4\tev$, then the decay length for the
NLSP is $c\tau\sim 10$~km for $\mcnone=100\gev$; while $\sqrt F\sim
100\tev$ implies a short but vertexable decay length.  A similar result
is obtained in the case of a charged NLSP.  Thus, if $\sqrt{F}$ is
sufficiently large, the charged NLSP will be semi-stable and may decay
outside of the collider detector.

Finally, if R-parity violation is present, the decay rate of the LSP in
SUGRA-based models (or the NLSP in R-parity-violating GMSB models) could
be in the relevant range
to yield visible secondary vertices.

\section{Supersymmetry searches at future colliders}

In this section, we consider the potential for discovering low-energy
supersymmetry at future colliders.  A variety of supersymmetric
signatures have been reviewed in Section 2, and we now apply
these to supersymmetry searches at future colliders.  Ideally,
experimental studies of supersymmetry should be as model-independent as
possible.  Ultimately, the goal of experimental studies of supersymmetry
is to measure as many of the 124 MSSM parameters (and any additional
parameters that can arise in non-minimal extensions) as possible.  In
practice, a fully general analysis will be difficult, particularly
during the initial supersymmetry discovery phase.  Thus, we focus the
discussion in this section on the expected phenomenology of
supersymmetry at the various future facilities under a number of
different model assumptions.  Eventually, if candidates for
supersymmetric phenomena are discovered, one would utilize
precision experimental measurements to map out the supersymmetric
parameter space and uncover the structure of the underlying
supersymmetry-breaking.

\subsection{SUGRA-based models}

We begin with the phenomenology of mSUGRA.  Of particular importance are
the relative sizes of the different supersymmetric particle masses.
Generic properties of the resulting superpartner mass spectrum are
discussed in \Ref{smartin}.  An important consequence of the mSUGRA mass
spectrum is that substantial phase space is available for most decays
occurring at each step in a given chain decay of a heavy
supersymmetric particle.


Extensive Monte Carlo studies have examined the region of mSUGRA
parameter space for which direct discovery of supersymmetric particles
at the Tevatron and the LHC will be possible \cite{baerreview}.  At the
hadron colliders, the ultimate supersymmetric mass reach is determined
by the searches for both the strongly-interacting superpartners (squarks and
gluinos) and the charginos/neutralinos.
Cascade decays of the produced squarks and gluinos lead
to events with jets, missing energy, and various numbers of leptons.
Pair production of charginos and/or neutralinos can produce
distinctive multi-lepton signatures.  The
chargino/neutralino searches primarily constrain the mSUGRA parameter
$m_{1/2}$, which can be translated into an equivalent bound on the
gluino mass.   As a result,
gluino and squark masses up to about 400 GeV can be probed at the
upcoming Tevatron Run-II; further improvements are projected at the
proposed TeV-33 upgrade \cite{tev33msugra}, where supersymmetric masses
up to about 600 GeV can be reached.  The maximum reach at the LHC is
generally
attained by searching for the $1\ell+{\rm jets}+\etmiss$ channel; one
will be able to discover squarks and gluinos with masses up to several
TeV \cite{bquarkbaer}. Some particularly important classes of events
include:
\begin{itemize}
\item
$pp\to \gl\gl\to \mbox{jets}+\etmiss$ and
$pp\to\gl\gl\to\ell^{\pm}\ell^{\pm}$+jets+$\etmiss$
(the like-sign dilepton signal \cite{likesign}).
The mass difference $\mgl-\mcpmone$ can be determined
from jet spectra end points,
while $\mcpmone-\mcnone$ can be roughly determined by analyzing various
distributions of kinematic observables in the like-sign
channel \cite{likesign,bquarkian,bartlsusy96}.
An absolute scale for $\mgl$ can be estimated (within an accuracy of
roughly $\pm 15\%$) by separating
the like-sign events into two hemispheres corresponding to the two $\gl$'s
\cite{likesign}, by a similar separation in the jets+$\etmiss$
channel \cite{bquarkbaer}, or variations thereof \cite{bquarkian,bartlsusy96}.
\item $pp\to \cpmone\cntwo\to(\ell^{\pm}\nu\cnone)(\ell^+\ell^-\cnone)$,
which yields a tri-lepton + $\etmiss$ final state.  The mass difference
$\mcntwo-\mcnone$ is easily determined~\footnote{In some cases,
$\mcntwo-\mcnone$ can still be determined if $\chinb$ is produced at some
step in a supersymmetric decay chain.}
if enough events are available \cite{bcpttri}.
\item $pp\to\slep\slep\to 2\ell+\etmiss$, detectable
at the LHC for $\mslep\lsim 300\gev$ \cite{bquarkbaer}.
\item Squarks will be pair produced and, for $m_0\gg\mhalf$,  would lead to
$\gl\gl$ events with two extra jets emerging from the
primary $\sq\to q\gl$ decays.
\end{itemize}

The LHC provides significant
opportunities for precision measurements of the mSUGRA parameters
\cite{bquarkian}.  In general, one expects large samples of
supersymmetric events with distinguishing features that allow an
efficient separation from Standard Model backgrounds.  The biggest
challenge in analyzing these events may be in distinguishing one set of
supersymmetric signals from another.  Within the mSUGRA framework, the
parameter space is small enough to permit the untangling of the various
signals and allows one to extract the mSUGRA parameters with some
precision.


Important discovery modes at the NLC include the following \cite{nlcreport}:
\begin{itemize}
\item $\epem\to \cpone\cmone\to (q\anti q\cnone~\mbox{or}~\ell\nu \cnone)+
(q\anti q\cnone~\mbox{or}~\ell\nu \cnone)$;
\item $\epem\to \slep^+\slep^-\to
(\ell^+\cnone~\mbox{or}~\anti{\nu}\cpone)+
(\ell^-\cnone~\mbox{or}~\nu \cmone)$.
\end{itemize}
In both cases, the masses of the initially produced supersymmetric
particles as well as the final state neutralinos and charginos
will be well-measured.  Here, one is able to make use of the
energy spectra end points and beam energy constraints to make precision
measurements of masses and determine the underlying supersymmetric
parameters.  Polarization of the beams is an essential tool that can be
used to enhance signals while suppressing Standard Model backgrounds.
Moreover, polarization can be employed to separate out various
supersymmetric contributions in order to explore the inherent chiral
structure of the interactions.
The supersymmetric mass reach is limited by the center-of-mass energy of
the NLC.  For example, if the scalar mass parameter $m_0$ is too large,
squark and slepton pair production will be kinematically forbidden.  To
probe values of $m_0\sim 1$---$1.5\tev$ requires a collider energy in
the range of $\rts\gsim 2$---$3\tev$. It could be that such energies
will be more easily achieved at a future $\mupmum$ collider.

The strength of the lepton colliders lies in the ability to analyze
supersymmetric signals and make precision measurements of observables.
Ideally, one would like to measure the underlying supersymmetric
parameters without prejudice.  One could then test the model
assumptions, and study possible deviations.  The most efficient way to
carry out such a program is to set the lepton collider center-of-mass
energy to the appropriate value of $\rts$ in order to first study the
light supersymmetric
spectrum (lightest charginos and neutralinos and sleptons).
In this way, one limits the interference among competing supersymmetric
signals.  Experimentation at the lepton colliders then can provide
model-independent measurements of the associated
underlying supersymmetric parameters.
Once these parameters are ascertained, one can analyze with
more confidence events with heavy supersymmetric particles decaying via
complex decay chains.  Thus, the NLC and LHC supersymmetric
searches are complementary.

Beyond mSUGRA, the MSSM parameter space becomes more complex.  It is
possible to perturb the mSUGRA model by adding some non-universality
among the scalar mass parameters without generating phenomenologically
unacceptable flavor changing neutral currents.
There has been no systematic analysis of the
resulting phenomenology at future colliders.  (The implications of
non-universal scalar masses for LHC phenomenology were briefly addressed
in \Ref{snowtheory2}.)  Nevertheless, the possible non-degeneracy of
squarks could have a significant impact on the search for squarks at
hadron colliders.  In particular, in mSUGRA models one typically finds
that four flavors of squarks (with two squark eigenstates per flavor)
and $\widetilde b_R$ are nearly mass-degenerate, while the masses of
$\widetilde b_L$ and the top-squark mass eigenstates could be
significantly different~\footnote{If $\tanb\gg 1$, then $\widetilde
b_L$--$\widetilde b_R$ mixing can be significant, in which case the two
bottom-squark mass eigenstates could also be significantly split in mass
from the first two generations of squarks.}.  This means that the {\it
observed} cross-section for the production of squark pairs at hadron
colliders would be enhanced by a multiplicity factor of eight or larger
(depending on the number of approximately mass-degenerate squark
species).  Clearly, if some of the first and second generation squarks
are split in mass, the relevant effective cross-sections are smaller.
This could lead to more background contamination of squark signals at
hadron colliders.  The impact of squark non-degeneracy on the discovery
mass reach for squarks at the Tevatron and LHC has not yet been
analyzed.

It is also possible to introduce arbitrary non-universal gaugino mass
parameters (at the high-energy scale).  For example, suppose that the
non-universal gaugino masses at the high-energy scale imply that the
gaugino mass parameters at the low-energy scale satisfy $M_2<M_1$, \ie,
the SU(2)-gaugino component is dominant in the lightest chargino and
neutralino \cite{guniondrees1,guniondrees2}.
In this case, the $\cnone$ and $\cpmone$ can be
closely degenerate, in which case the visible decay products in
$\cpone\to \cnone+X$ decays will be very soft and difficult to detect.
Consequences for chargino and neutralino detection in $\epem$ and
$\mupmum$ collisions, including the importance of the $\epem\to
\gam\cpone\cmone$ production channel, are discussed in
Refs.~\cite{guniondrees1,guniondrees2}.
There is also the possibility that $\mgl\sim \mcpmone\simeq\mcnone$.
The decay products in the $\gl$ decay chain would then be very soft,
and isolation of $\gl\gl$ events would be much more difficult at hadron
colliders than in the usual mSUGRA case. In particular, hard jets
in association with missing energy would be much
rarer, since they would only arise from initial state radiation.
The corresponding reduction in supersymmetric parameter space
coverage at the Tevatron Main Injector is explored in
Ref.~\cite{guniondrees2}.

As a second example, consider the case where the low-energy gaugino mass
parameters satisfy $M_2\sim M_1$~\footnote{We remind the reader that
gaugino mass unification at the high-energy scale would predict
$M_2\simeq 2M_1$.}.  If we also assume
that $\tanb\sim 1$ and $|\mu|<M_1$, $M_2$~\footnote{To achieve
such a small $\mu$-parameter requires, \eg,  some non-universality among
scalar masses of the form $m_{H_1}^2\neq m_{\sq}^2$, $m_{\slep}^2$.},
then the lightest two neutralinos are nearly a pure photino and
higgsino respectively, \ie,
$\chinb\simeq\wt\gamma$ and $\china\simeq\wt H$.
For this choice of MSSM parameters, one finds that the rate for
the one-loop decay
$\cntwo\to \gam \cnone$ dominates over all tree level decays of $\chinb$
and $\br(\sel\to e\cntwo)\gg\br(\sel\to e\cnone)$.
Clearly, the resulting phenomenology \cite{eegamgamkane}
differs substantially from mSUGRA expectations.  This scenario was
inspired by the CDF $ee\gam\gam$ event \cite{eegamgam}.
Suppose that the $ee\gam\gam$ event resulted from $\sel\sel$ production,
where $\sel\to e\chinb\to e\gamma\china$.
Then in the model of \Ref{eegamgamkane},
one would expect a number of other distinctive supersymmetric
signals to be observable at LEP-2 (running at its maximal energy) and at
Run-II of the Tevatron.  In particular, LEP-2 would expect events of the
type: $\ell\ell+X+\etmiss$ and $\gam\gam+X+\etmiss$, while Tevatron
would expect events of the type:
$\ell\ell+X+\etmiss$, $\gam\gam+X+\etmiss$,
$\ell\gam+X+\etmiss$, $\ell\ell\gam+X+\etmiss$,
$\ell\gam\gam+X+\etmiss$, and $\ell\ell\ell+X+\etmiss$.
In the above signatures, $X$ stands for
additional leptons, photons, and/or jets.
These signatures can also arise in GMSB models, although the kinematics
of the various events can often be distinguished.

\subsection{GMSB-based models}

The collider signals for GMSB models depend critically on the NLSP
identity and its lifetime (or equivalently, its decay length).  Thus, we
examine the phenomenology of both promptly-decaying and longer-lived
NLSP's.  In the latter case, the number of decays where one or both
NLSP's decay within a radial distance $R$ is proportional to
$[1-\exp(-2R/c\tau)]\simeq 2R/(c\tau)$.  For large $c\tau$, most decays
would be non-prompt, with many occurring in the outer parts of the
detector or completely outside the detector. To maximize sensitivity to
GMSB models and fully cover the $(\sqrt F,\Lambda)$ parameter space, we
must develop strategies to detect decays that are delayed, but not
necessarily so delayed as to be beyond current detector coverage and/or
specialized extensions of current detectors.

In the discussion below, we focus on various cases, where the NLSP is a
neutralino dominated by its U(1)-gaugino ($\widetilde B$) or Higgsino
($\widetilde H$) components, and where the NLSP is the lightest charged
slepton (usually the $\staur$).  We first address the case of prompt
decays, and then indicate the appropriate strategies for the case of the
longer-lived NLSP.

\begin{itemize}
\item {{\bf Promptly-decaying NLSP:} $\bold{\cnone\simeq\wtil B}$}
\end{itemize}

We focus on the production of the neutralinos, charginos, and sleptons
since these are the lightest of the supersymmetric
particles in the GMSB models.  The possible decays of the NLSP in this
case are: $\wt B\to \gamma\gtino$ or $\wt B\to  Z\gtino$.  The
latter is only relevant for the case of a heavier NLSP (and
moreover is suppressed by $\tan^2\theta_W$). It will be
ignored in the following discussion.

At hadronic colliders, the $\cnone\cnone$ production rate is small, but
rates for $\cpone\cmone\to W^{(\star)}W^{(\star)}\cnone\cnone\to
W^{(\star)}W^{(\star)} \gam\gam+\etmiss$, $\slepr\slepr\to
\ell^+\ell^-\cnone\cnone\to\ell^+\ell^-\gam\gam+\etmiss$,
$\slepl\slepl\to\ell^+\ell^-\cnone\cnone\to
\ell^+\ell^-\gam\gam+\etmiss$, \etc\ will all be substantial.
Implications for GMSB phenomenology at the Tevatron can be found in
Refs.~\cite{gmsb,baerbrhliktata,smartingmsbsearch}.  It is possible to
envision GMSB parameters such that the $ee\gam\gam+\etmiss$ CDF
event \cite{eegamgam}
corresponds to selectron pair production followed by $\widetilde e\to
e\china$ with $\china\to\gamma\gtino$
\cite{gmsb,eegamgamkane,smartingmsbsearch,lopezgmsb}.  However, in this
region of GMSB parameter space, other supersymmetric signals should be
prevalent, such as $\snul\slepl\to\ell\gam\gam+\etmiss$ and
$\snul\snul\to\gam\gam+\etmiss$.  The $\cntwo\cpmone$ and $\cpone\cmone$
rates would also be significant and lead to $X\gam\gam+\etmiss$ with
$X=\ell^\pm,\ell^+\ell^{\prime\,-},\ell^+\ell^-\ell^{\prime\,\pm}$.
Limits on these event rates from current CDF and D0 data already
eliminate much, if not all, of the parameter space that could lead to
the CDF $ee\gam\gam$ event \cite{limitsoneegg}.

At LEP-2/NLC \cite{lepnlcgmsbgam}, the rate for the simplest signal,
$\epem\to \cnone \gtino \to \gam+\etmiss$, is expected to be very
small.  A more robust
channel is $\epem\to \cnone\cnone\to \gam\gam+\etmiss$ with
a (flat) spectrum of photon energies in the range ${1\over
4}\rts(1-\beta)\leq E_\gam\leq {1\over 4}\rts(1+\beta)$.

\begin{itemize}
\item {{\bf Promptly decaying NLSP:} $\bold{\cnone\simeq\wtil H}$}
\end{itemize}

The possible decays of the NLSP in this case are: $\wtil H\to \gtino
+\hl,\hh,\ha$, depending on the Higgs masses.  If the corresponding
two-body decays are not kinematically possible, then
three-body decays (where the corresponding Higgs state is virtual) may
become relevant.  However, in realistic cases, one expects $\cnone$ to
contain small but non-negligible gaugino components, in which case
the rate for
$\cnone\to\gtino\gamma$ would dominate all three-body decays.
In what follows, we assume that the two-body decay $\wtil
H\to\gtino\hl$ is kinematically allowed and dominant.
The supersymmetric signals that would emerge
at both Tevatron/LHC and LEP-2/NLC would then be
$4b+ X + \etmiss$ final states, where $X$ represents the
decay products emerging from the cascade chain decays
of the more massive supersymmetric particles.
Of course, at LEP-2/NLC direct
production of higgsino pairs, $e^+e^-\to \wt H\wt H$ (via virtual
$s$-channel $Z$-exchange) would be possible in general, leading to
pure $4b+\etmiss$ final states.

\begin{itemize}
\item {{\bf Promptly decaying NLSP:} $\bold{\slepr}$}
\end{itemize}

The dominant slepton decay modes
are: $\slepr^\pm\to \ell^\pm \gtino$ and
$\slepl^\pm\to\ell^\pm \cnonestar\to
\ell^\pm(\slepr^\pm\ell^\mp)^\prime
\to \ell^\pm(\ell^\pm\ell^\mp)^\prime \gtino$.
The $\cnone$ will first decay to $\ell \slepl$ and $\ell\slepr$, followed
by the above decays.

At both the Tevatron/LHC and LEP-2/NLC, typical pair production events
will end with $\slepr\slepr\to \ell^+\ell^-+\etmiss$, generally in
association with a variety
of cascade chain decay products. The lepton energy spectrum
will be flat in the $\slepr\slepr$ center of mass.
Of course, pure $\slepr\slepr$ production is possible
at LEP/NLC and the $\slepr\slepr$ center of mass would be
the same as the $\epem$ center of mass. Other simple signals at LEP/NLC,
would include $\slepl\slepl\to 6\ell +\etmiss$.

If a slepton is the NLSP, it is most likely to be the $\staur$.  If this
state is sufficiently lighter than the $\wt e_R$ and $\wt \mu_R$, then
$\wt e_R \to e \staur\tau$ and $\wtil \mu_R\to \mu \staur\tau$ decays
(via the $\wtil B$ component of the mediating virtual neutralino) might
dominate over the direct $\wt e_R\to e\gtino$ and $\wt \mu_R\to
\mu\gtino$ decays, and all final states would cascade to $\tau$'s. The
relative importance of these different possible decays has been examined
in Ref.~\cite{conlsp}. A study of this scenario at LEP-2 has been
performed in Ref.~\cite{dicusstau}.

\begin{itemize}
\item {{\bf Longer-lived NLSP:} $\bold{\slepr}$}
\end{itemize}

If the $\slepr$ mainly decays before reaching the electromagnetic
calorimeter, then one should look for a charged lepton that suddenly
appears a finite distance from the interaction region, with non-zero
impact parameter as measured by either the vertex detector or the
electromagnetic calorimeter. Leading up to this decay would be a heavily
ionizing track with $\beta<1$ (as could be measured if a magnetic field
is present).

If the $\slepr$ reaches the electromagnetic and hadronic calorimeters,
then it behaves much like a heavy muon, presumably interacting in the
muon chambers or exiting the detector if it does not decay first.
Limits on such objects should be pursued.
There will be many sources of $\slepr$ production, including direct
slepton pair production, and cascade decays resulting from the
production of gluinos, squarks, and charginos \cite{chengunionslep}.
Based on current Tevatron data, a charged pseudo-stable
$\slepr$ can be ruled out with a mass up to about $80$--$100\gev$.
Similar limits can probably be extracted from LEP-2 data.

\begin{itemize}
\item {{\bf Longer-lived NLSP:} $\bold{\china}$}
\end{itemize}

This is a much more difficult case. As before, we assume that the
dominant decay of the NLSP in this case is $\china\to\gamma\gtino$.
Clearly, the sensitivity of detectors to delayed $\gam$ appearance
signals will be of great importance.  If the $\china$ escapes the
detector before decaying, then the corresponding missing energy
signatures are the same as those occurring in SUGRA-based models.

At the Tevatron, standard supersymmetry signals (\eg, jets or
tri-leptons plus $\etmiss$) are viable if $\Lambda\lsim 30$--$70\tev$
(given an integrated luminosity of
$L=0.1$--$30\fbi$) independent of the magnitude of $\sqrt F$
\cite{chenguniongampub,chenguniongamprelim}. Meanwhile, the prompt
$\cnone\to\gam\gtino$ decay signals discussed earlier are viable only in
a region defined by $\sqrt F\lsim 500\tev$ at low $\Lambda$, rising to
$\sqrt F\lsim 1000\tev$ at $\Lambda\sim
120\tev$~\cite{chenguniongampub,chenguniongamprelim}.
This leaves a significant region of
$(\sqrt{F},\Lambda)$ parameter space that can only be probed by the
delayed $\cnone\to \gam\gtino$ decays
\cite{chenguniongampub,chenguniongamprelim}.

The ability to search for delayed-decay signals
is rather critically dependent upon the detector
design. The possible signals
include the following \cite{chenguniongampub,chenguniongamprelim}:
(i) looking for isolated energy deposits
(due to the $\gam$ from the $\cnone$ decay)
in the outer hadronic calorimeter cells of the D0 detector;
(ii) searching for events where the delayed-decay
photon is identified by a large (transverse)
impact parameter as it passes into the electromagnetic
calorimeter; and (iii)
looking for delayed decays where the photon first emerges outside
the main detector and is instead observed in a scintillator array (or similar
device) placed at a substantial distance from the detector.
The observed signal will always contain missing energy from one
or more emitted gravitinos and/or from $\cnone$'s that do not decay
inside the detector.  Thus, by requiring large missing energy, the
backgrounds can be greatly reduced while maintaining good efficiency
for the GMSB signal. In combination, the above techniques
may~\footnote{Event rates are significant even after very strong cuts
on jets, photon energy and missing energy, but detailed background
calculations remain to be done.}
allow the detection of supersymmetric particle production at the
Tevatron in the GMSB parameter region
$\sqrt F\lsim 3000\tev$ and $\Lambda\lsim 150\tev$.

\subsection{R-parity violating (RPV) models}

In R-parity violating models, the LSP is no longer stable~\footnote{We
assume that the gravitino is not relevant for RPV phenomenology, as in
SUGRA-based models.}.
The relevant signals depend upon the nature of the LSP decay.  The
phenomenology depends on which R-parity violating couplings [\Eq{rpv}]
are present.  Only a brief discussion will be given here; for further
details, see \Ref{dreiner}.


At the Tevatron and LHC \cite{likesignrviol},
consider $\wt g\wt g$ production followed by gluino decay via the usual
set of possible decay chains ending up with the LSP plus Standard Model
particles. Until this point, all decays have involved only R-parity
conserving interactions~\footnote{By assumption, the strengths of
the R-parity conserving interactions
are significantly larger than the corresponding RPV-interaction
strengths.}. The
RPV-interactions now enter in the decay of the LSP.  We shall assume in
the following discussion that the $\china$ is the LSP, although other
possible choices can also be considered.

If $\lambda_B\neq 0$, then the dominant decay of $\cnone$ would result
in the production of a three-jet final state
($\cnone\to jjj$). The large jet backgrounds
imply that we would need to rely on the like-sign dilepton signal
(which would still be viable despite the absence of missing energy
in the events).  In general,
this signal turns out to be sufficient for supersymmetry discovery
out to gluino masses somewhat above 1 TeV.
However, if the leptons of the like-sign dilepton signal are very soft,
then the discovery reach would be much reduced~\footnote{Soft leptons
would occur in models where $\mcpmone\sim\mcnone$, which requires
non-universal gaugino masses \cite{guniondrees1,snowtheory2}.}.
This is one of the few cases where one could miss discovering low-energy
supersymmetry at the LHC.
If $\lambda_L$ dominates $\cnone$ decays,
$\cnone\to \mu^\pm e^\mp\nu,e^\pm e^\mp \nu$, and there would be many
very distinctive multi-lepton signals.
If $\lambda_L^\prime$ is dominant, then
$\cnone\to \ell jj$ and again there would
be distinctive multi-lepton signals.

More generally, many normally invisible events become visible.
An important example is sneutrino pair production.
Even if the dominant decay of the sneutrino is
$\snu\to\nu\cnone$ (which is likely if $\msnu>\mcnone$),
a visible signal emerges from the $\cnone$ decay as sketched above.
Of course, for large enough $\lambda_L$ or $\lambda_L^\prime$
the $\snu$'s would have significant branching ratio for
decay to charged lepton pairs or jet pairs, respectively.
Indeed, such decays might dominate if $\msnu<\mcnone$.


At LEP-2, NLC or the muon collider \cite{eerpv,schannel,fghsnu},
the simplest process is
\beq
\epem\to\cnone\cnone\to \underbrace{(jjj)(jjj)}_{\lambda_B},
~~\underbrace{(\ell\ell\nu)(\ell\ell\nu)}_{\lambda_L},
~~\underbrace{(\ell jj)(\ell jj)}_{\lambda_L^\prime}
\label{simppro}
\eeq
(or the $\mupmum$ collision analogue),
where the relevant RPV-coupling is indicated below the corresponding
signal. Substantial rates for
equally distinctive signals from production of more massive
supersymmetric particles (including sneutrino pair production)
would also be present.  All these processes (if kinematically allowed)
should yield observable supersymmetric signals.  Some limits from LEP data
already exist \cite{lepnlcrpv}.
Of particular potential importance for non-zero $\lambda_L$ is
\hbox{$s$-channel}
resonant production of a sneutrino in $e^+e^-$ \cite{schannel}
and $\mupmum$ \cite{fghsnu} collisions.
In particular, at $\mupmum$ colliders this process is detectable down
to quite small values of the appropriate $\lambda_L$, and could be
of great importance as a means of actually determining the
R-parity-violating couplings. Indeed, for small R-parity-violating
couplings, absolute measurements of the couplings
through other processes are extremely difficult. This
is because such a measurement would typically require the R-parity-violating
effects to be competitive with an R-parity-conserving process of known
interaction strength. (For example, R-parity-violating
neutralino branching ratios constrain only ratios of the R-parity-violating
couplings.) Since sneutrino pair production would have been observed
at the LHC, NLC and/or the muon collider, it would be easy
to center on the sneutrino resonance in order to perform the crucial
sneutrino factory measurements.

We end this section with two additional remarks.  First, if the RPV
coupling strengths are very small, then the RPV-violating decay of the
LSP (\eg, $\cnone$) could occur a substantial distance from the primary
interaction point, but still within the detector (or at least not far
outside the detector). The general techniques for detecting such delayed
decays outlined at the end of Section~3.2 would again be relevant.  It
is particularly important to note that observation of the delayed decays
would allow a determination of the absolute strengths of the RPV
couplings.  Second, one should not neglect the possibility that RPV
couplings could be present in GMSB models.  If the RPV couplings
are substantial, then the RPV decays of the NLSP will dominate its
R-parity-conserving decays into $\gtino+X$~\cite{gmsbrpv},
and all the RPV phenomenology described in this section
will apply. For smaller RPV couplings, there could be competition
between the RPV
decays and the $\gtino+X$ decays of the NLSP.

\section{Summary and Conclusions}

Much effort has been directed at trying to develop strategies
for precision measurements to establish the underlying supersymmetric
structure of the interactions and to distinguish among models.  However,
we are far from understanding all possible facets of
the most general MSSM parameter
space (even restricted to those regions that are phenomenologically
viable).  Moreover, the phenomenology of non-minimal and alternative
low-energy supersymmetric models (such as models with R-parity
violation) and the consequences for collider physics have only recently
begun to attract significant attention.  The variety of possible
non-minimal models of low-energy supersymmetry presents an additional
challenge to experimenters who plan on searching for supersymmetry at
future colliders.

If supersymmetry is discovered, it will provide a plethora of
experimental signals and theoretical analyses.
The many phenomenological manifestations
and parameters of supersymmetry suggest that many years of experimental
work will be required before it will be possible to determine
the precise nature of supersymmetry-breaking and its
implications for a more fundamental theory of particle interactions.

\section*{Acknowledgements}

This work was supported in part by the Department of Energy.
JFG is supported in part by
the Davis Institute for High Energy Physics.


\section*{References}

\begin{thebibliography}{99}

\bibitem{Haber85}
H.E.~Haber and G.L.~Kane, {\sl Phys. Rep.} {\bf 117} (1985) 75.

\bibitem{smartin}
S.P.~Martin, hep-ph/9709356, chapter in this volume.

\bibitem{sutter}
S. Dimopoulos and D. Sutter,  {\sl Nucl. Phys.} {\bf B452} (1995) 496;
D.W. Sutter, Stanford Ph.~D. thesis, hep-ph/9704390.

\bibitem{habersusy97}
H.E. Haber, {\sl Nucl. Phys. B (Proc. Suppl.)} {\bf 62A-C} (1998) 469.

\bibitem{sugra}
H.P. Nilles, {\sl Phys. Rep.} {\bf 110} (1984) 1;
P. Nath, R. Arnowitt, and A.H. Chamseddine, {\it Applied
$N=1$ Supergravity} (World Scientific, Singapore, 1984).


\bibitem{giudice}
For a review of gauge-mediated supersymmetry breaking, see G.F. Giudice
and R. Rattazzi, CERN-TH/97-380 [hep-ph/9801271], submitted to {\sl
Physics Reports}; and chapter in this volume


\bibitem{dreiner}
For a recent review and guide to the literature, see H. Dreiner,
hep-ph/9707435, chapter in this volume.

\bibitem{leveille}
G.L. Kane and J.P. Leveille, {\sl Phys. Lett.} {\bf 112B} (1982) 227.

\bibitem{chain}
H. Baer, J. Ellis, G. Gelmini, D. Nanopoulos and X. Tata, {\sl
Phys. Lett.} {\bf 161B} (1985) 175; G. Gamberini, \ZPC 30 605 1986 ;
H. Baer, V. Barger, D. Karatas and X. Tata, {\sl Phys. Rev.}
{\bf D36} (1987) 96;
R.M. Barnett, J.F. Gunion and H.E. Haber, \PRL 60 401 1988 ; \PRD 37
1892 1988 ;

\bibitem{enhancedbs}
H. Baer, X. Tata and J. Woodside, \PRD 41 906 1990 ;
\PRD 45 142 1992 .

\bibitem{guniondrees1}
C.H. Chen, M. Drees, and J.F. Gunion,  \PRL 76 2002 1996 .

\bibitem{guniondrees2}
C.H. Chen, M. Drees, and J.F. Gunion,  \PRD 55 330 1997 .

\bibitem{glennys}
G.R. Farrar, {\sl Phys. Rev. Lett.} {\bf 76} (1996) 4111, 4115; \PRD
51 3904 1995 .  For a recent review, see G.R. Farrar, 
{\sl Nucl. Phys. B (Proc. Suppl.)} {\bf 62A-C} (1998) 485.

\bibitem{nogluinos}
F. Csikor and Z. Fodor, {\sl Phys. Rev. Lett.} {\bf 78} (1997) 4335;
preprint ITP-BUDAPEST-538 [hep-ph/9712269];
Z. Nagy and Z. Trocsanyi, hep-ph/9708343; hep-ph/9712385;
J. Adams {\it et al.} [KTeV Collaboration],
{\sl Phys. Rev. Lett.} {\bf 79} (1997) 4083;
P. Abreu {\it et al.} [DELPHI Collaboration],
{\sl Phys. Lett.} {\bf B414} (1997) 401;
R. Barate {\it et al.} [ALEPH Collaboration], \ZPC 96 1 1997 .


\bibitem{gunbarcelona} J.F. Gunion, to appear in Proceedings of the
{\it
International Workshop on Quantum Effects in the MSSM}, UAB, Barcelona,
9--13 September 1997, edited by J. Sola (World Scientific, Singapore).

\bibitem{raby}
S. Raby, \PRD 56 2852 1997 ; preprint OHSTPY-HEP-T-97-024 [hep-ph/9712254].

\bibitem{conlsp}
S. Ambrosanio, G.D. Kribs, and S.P. Martin, SLAC-PUB-7668 (1997)
[hep-ph/9710217].


\bibitem{likesign} R.M. Barnett, J.F. Gunion and H.E. Haber, \PLB 315 349
1993 .

\bibitem{bcpttri} For an LHC study, see H. Baer, C.H. Chen, F. Paige and
X. Tata, \PRD 50 4508 1994 . Tevatron studies by theorists include:
H. Baer and X. Tata, \PRD 47 2739 1993 ; H. Baer, C. Kao and X. Tata,
\PRD 48 5175 1993 ;
H. Baer, C.-H. Chen, C. Kao and X. Tata,
\PRD 52 1565 1995 ; S. Ambrosanio, G.L. Kane, G.D. Kribs, S.P. Martin
and S. Mrenna, \PRD 54 5395 1996 .
The most recent supersymmetric limits based on
Tevatron tri-lepton searches are given in B. Abbott \etal\ [D0
Collaboration], FERMILAB-PUB-97-153-E (1997) [hep-ex/9705015]; and,
F. Abe \etal\ [CDF Collaboration], \PRL 76 4307 1996 .

\bibitem{bquarkbaer}
H. Baer, C.-H. Chen, F. Paige and X. Tata, \PRD 52 2746 1995 ;
\PRD 53 6241 1996 .

\bibitem{bquarkian} I. Hinchliffe, F.E. Paige, M.D. Shapiro,
J. Soderqvist and W. Yao, \PRD 55 5520 1997 .

\bibitem{snowtheory2} G. Anderson, C.H. Chen, J.F. Gunion, J. Lykken,
T. Moroi, Y. Yamada, in {\it New Directions for High-Energy Physics},
Proceedings of the 1996 DPF/DPB Summer Study on High Energy Physics,
Snowmass '96, edited by D.G. Cassel, L.T. Gennari and R.H. Siemann
(Stanford Linear Accelerator Center, Stanford, CA, 1997) pp.~669--673.

\bibitem{wyler} H. Komatsu and J. Kubo, {\sl Phys. Lett.} {\sl 157B}
(1985) 90; \NPB 263 265 1986 ; H.E. Haber, G.L. Kane and M. Quiros, {\sl
Phys. Lett.} {\sl 160B} (1985) 297; \NPB 273 333 1986 ; R. Barbieri,
G. Gamberini, G.F. Giudice and G. Ridolfi, \NPB 296 75 1988 ; H.E. Haber
and D. Wyler, \NPB 323 267 1989 .

\bibitem{eegamgamkane}
S. Ambrosanio, G.L. Kane, G.D. Kribs, S.P. Martin
and S. Mrenna, \PRD 55 1372 1997 .


\bibitem{baerreview}
H. Baer {\it et al.}, in {\it Electroweak Symmetry Breaking and New
Physics at the TeV Scale}, edited by T.L. Barklow, S. Dawson, H.E.
Haber and J.L. Siegrist (World Scientific, Singapore, 1996) pp.~216--291.

\bibitem{tev33msugra} See, \eg, H. Baer. C.-H. Chen, F.
Paige and X. Tata, \PRD 54 5866 1996 ; D. Amidei \etal\ [TeV-2000 Study
Group], FERMILAB-PUB-96-082; and references therein.


\bibitem{bartlsusy96} A. Bartl \etal, in {\it New Directions for High-Energy
Physics}, Proceedings of the 1996 DPF/DPB Summer Study on High
Energy Physics, Snowmass '96, edited by D.G. Cassel, L.T. Gennari
and R.H. Siemann
(Stanford Linear Accelerator Center, Stanford, CA, 1997) pp.~693--707.

\bibitem{nlcreport} T. Tsukamoto, K. Fujii, H. Murayama, M. Yamaguchi
and Y. Okada, \PRD 51 3153 1995 ; J.L. Feng, M.E. Peskin, H. Murayama
and X. Tata, {\sl Phys. Rev.} {\bf D52} (1995) 1418; H. Baer, R. Munroe
and X. Tata, \PRD 54 6735 1996 ; J.L. Feng and M.J. Strassler, {\sl
Phys. Rev.} {\bf D55} (1997) 1326; H. Murayama and M.E. Peskin, {\sl
Ann. Rev. Nucl. Part. Sci.} {\bf 46} (1996) 533.


\bibitem{eegamgam}
F. Abe \etal\ [CDF Collaboration], FERMILAB-PUB-98-024-E [hep-ex/9801019].

\bibitem{gmsb}
S. Dimopoulos, S. Thomas and J.D. Wells, \PRD 54 3283 1996 ; \NPB 488 39 1997 .


\bibitem{baerbrhliktata}
H. Baer, M. Brhlik, C.-H. Chen and X. Tata, \PRD 55 4463 1997 .

\bibitem{smartingmsbsearch}
S. Ambrosanio, G.L. Kane, G.D. Kribs, S.P. Martin, \PRD 54 5395 1996 .


\bibitem{lopezgmsb} S. Dimopoulos, M. Dine, S. Raby and S. Thomas, \PRL
76 3494 1996 ; S. Ambrosanio, G.L. Kane, G.D. Kribs, S.P. Martin and
S. Mrenna, \PRL 76 3498 1996 .


\bibitem{limitsoneegg}
B. Abbott \etal\ [D0 Collaboration],
{\sl Phys. Rev. Lett.} {\bf 80} (1998) 442;
E. Flattum [for the CDF and D0 Collaborations], FERMILAB-Conf-97/404-E.

\bibitem{lepnlcgmsbgam} J. Lopez, D. Nanopoulos and A. Zichichi, \PRD 55
5813 1997 ; \PRL 77 5168 1996 ; S. Dimopoulos, M. Dine, S. Raby and
S. Thomas, \PRL 76 3494 1996 ; S. Ambrosanio, G.D. Kribs and
S.P. Martin, \PRD 56 1761 1997 ; D.R. Stump, M. Wiest and C.-P. Yuan,
\PRD 54 1936 1996 ; J. Bagger, K. Matchev, D. Pierce and R.-J. Zhang,
\PRL 78 1002 1997 .

\bibitem{dicusstau}
D.A. Dicus, B. Dutta and S. Nandi, \PRL 78 3055 1997 .

\bibitem{chengunionslep}
J.L. Feng and T. Moroi,  LBNL-41133 [hep-ph/9712499];
C.-H. Chen and J.F. Gunion, work in progress.

\bibitem{chenguniongampub}
C.-H. Chen and J.F. Gunion, UCD-97-15 (1997) [hep-ph/9707302].

\bibitem{chenguniongamprelim}
C.-H. Chen and J.F. Gunion, UCD-98-3 [hep-ph/9802252].

\bibitem{likesignrviol} P. Binetruy and J.F. Gunion, in {\it Heavy
Flavors and High Energy Collisions in the 1---100~TeV Range},
Proceedings of the INFN Eloisatron Project Workshop, Erice, Italy, June
10--27, 1988, edited by A. Ali and L. Cifarelli (Plenum Press, New York,
1989) p.~489; H. Dreiner and G.G. Ross, {\sl Nucl. Phys.} {\bf B365}
(1991) 597; H. Dreiner, M. Guchait and D.P. Roy, \PRD 49 3270 1994 ;
V. Barger, M.S. Berger, P. Ohmann, R.J.N. Phillips, \PRD 50 4299 1994 ;
H. Baer, C. Kao and X. Tata, \PRD 51 2180 1995 ; H. Baer, C.-H. Chen and
X. Tata, \PRD 55 1466 1997 ; A. Bartl \etal, \NPB 502 19 1997 .



\bibitem{eerpv}
G. Bhattacharyya, D. Choudhury and K. Sridhar, {\sl Phys. Lett.}
{\bf B355} (1995) 193;
G. Bhattacharyya, J. Ellis and K. Sridhar, {\sl Mod. Phys. Lett.}
{\bf A10} (1995) 1583;
J.C. Romao, F. de Campos, M.A. Garcia-Jareno, M.B. Magro, and J.W.F.
Valle, {\sl Nucl. Phys.} {\bf B482} (1996) 3;
K. Huitu, J. Maalampi and K. Puolamaki, preprint HIP-1997-24-TH
[hep-ph/9705406];
F. de Campos, O.J.P. Eboli, M.A. Garcia-Jareno, and J.W.F. Valle,
preprint IFUSP-1278 [hep-ph/9710545].

\bibitem{schannel} S. Dimopoulos and L.J. Hall, {\sl Phys. Lett.} {\bf
B207} (1988) 210; V. Barger, G.F. Giudice and T. Han, {\sl Phys. Rev.}
{\bf D40} (1989) 2987; R.M. Godbole, P. Roy and X. Tata, {\sl
Nucl. Phys.} {\bf B401} (1993) 67; J. Erler, J.L. Feng and N. Polonsky,
{\sl Phys. Rev. Lett.} {\bf 78} (1997) 3063; J. Kalinowski, R. Ruckl,
H. Spiesberger and P.M. Zerwas, {\sl Phys. Lett.} {\bf B406} (1997) 314.

\bibitem{fghsnu}
J.L. Feng, J.F. Gunion and T. Han, UCD-97-25 (1997) [hep-ph/9711414].


\bibitem{lepnlcrpv}
See, \eg, D. Buskuli \etal\ [ALEPH Collaboration], \PLB 384 461 1996 .

\bibitem{gmsbrpv} M. Carena, S. Pokorski and C.E.M. Wagner,
CERN-TH-97-373 [hep-ph/9801251].


\end{thebibliography}
\end{document}